# Drinkers' Intervention to Prevent Tuberculosis (DIPT Study)

**A study of the:** Makerere University – University of California, San Francisco (MU-UCSF) Research Collaboration and Mbarara University of Science and Technology (MUST) – University of California, San Francisco

**Sponsored by**: National Institute of Alcohol Abuse and Alcoholism (NIAAA)

**Principal Investigators:** Judith A. Hahn, PhD MA

University of California, San Francisco

Gabriel Chamie, MD, MPH

University of California, San Francisco

**Author:** Sara Lodi, PhD (biostatistician)

Boston University School of Public Health

## Introduction

The Drinkers' Intervention to Prevent TB (DIPT) is a randomized controlled trial designed to determine if incentive-based approaches can reduce alcohol use and improve medication adherence to isoniazid (INH) preventive therapy in persons with HIV (PWH) co-infected with tuberculosis (TB) who engage in heavy drinking in Uganda.

This statistical analysis plan (SAP) provides a detailed descriptions of the primary and secondary outcomes in the study and the corresponding statistical analyses.

## Study design

DIPT is a randomized, 2x2 factorial trial among adults co-infected with HIV/TB and who are heavy alcohol users in southwestern Uganda. The study is conducted at participating clinics in an urban setting (Mbarara city) and two rural settings (Rugazi and Ruhoko).

Figure 1 summarizes the study design of the trial. All eligible participants receive 6 months of INH and are randomly allocated to one of the following four arms: Arm 1: no incentives (control); Arm 2: financial incentives contingent on low alcohol use at INH refill visits; Arm 3: financial incentives contingent on high INH adherence at INH refill visits; Arm 4: financial incentives contingent on low alcohol use and on high INH adherence (rewarded independently). Low alcohol use at INH refill visits is measured via a negative POC ethyl glucuronide (EtG) urine test, defined as EtG<300 ng/mL. EtG is a metabolite of alcohol use that can be detected in urine for up to two to three days after heavy drinking.  Adherence to INH is measured via a positive POC IsoScreen test, a marker for INH ingestion in the prior 24 hours.

*Figure 1. Schematic of study design and aims of the Drinkers' Intervention to Prevent TB (DIPT) trial*

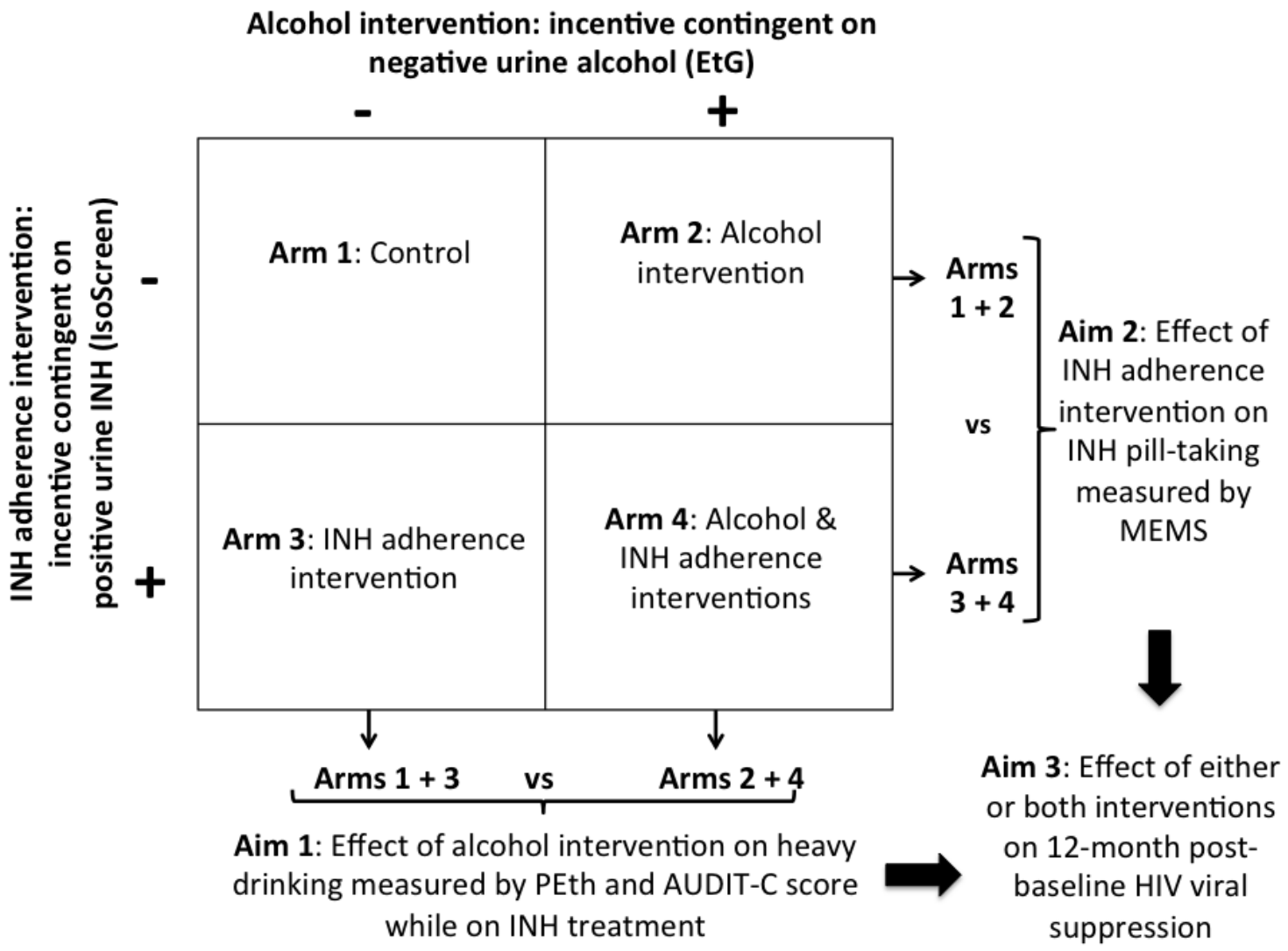

**Objectives**

The DIPT trial is designed to achieve three aims. Aim 1: To determine the effectiveness of financial incentives contingent on low alcohol use (Arms 2 and 4) versus no alcohol incentives (Arms 1 and 3) to reduce heavy drinking during INH preventive therapy. The hypothesis for Aim 1 is that participants receiving incentives for negative urine EtG testing will have lower alcohol use, determined via a medium-term alcohol biomarker during INH, compared to participants receiving no alcohol incentives. Aim 2: To determine the effectiveness of financial incentives contingent on high INH adherence (Arms 3 and 4) versus no INH adherence incentives (Arms 1 and 2) to increase INH adherence during INH preventive therapy. The hypothesis for Aim 2 is that participants receiving incentives for positive POC IsoScreen testing will have better adherence to INH, determined via objective monitoring, than participants receiving no INH adherence incentives. Aim 3: To determine the effectiveness of financial incentives (Arms 2, 3, and 4) versus no financial incentives (Arm 1) on HIV virological suppression after completing their INH preventive therapy. The hypothesis for Aim 3 is that participants receiving financial incentives have higher rates of HIV virological suppression six months after INH completion.

**Primary outcomes**

The study has two primary outcomes. The primary outcome for Aim 1 (i.e., to estimate the effect of incentives contingent on negative urine EtG tests) is no heavy alcohol consumption over the six months of INH preventive therapy, a binary variable. This is defined as a composite outcome of both PEth <35 ng/mL and negative AUDIT-C score at both the 3- and 6-month visits. We will use standard AUDIT-C cutoffs (≥3 for women and ≥4 for men) to define heavy drinking.

The primary outcome for Aim 2 (i.e., to estimate the effect of incentives contingent on a positive IsoScreen test) will be INH adherence measured as >90% pill-taking days by MEMS cap opening during the six-month course of INH or the time prescribed for those who were instructed to discontinue INH, a binary variable. The participants were given up to 9 months to complete a 6-month (180-pill) INH preventive therapy course. Therefore, MEMS adherence is defined as the number of pill bottle openings (no more than 1 per day counted) within a 270-day window divided by the number of prescribed doses (180, unless the participant discontinued the medication).

The outcome for Aim 3 is treated as a secondary outcome (see below).

**Secondary outcomes**

The secondary outcome for Aim 1 is PEth as a continuous variable measured at the 6-month visit.

The secondary outcome for Aim 2 is MEMS-measured adherence as a continuous variable (defined as the proportion of days with a bottle opening out of number of days prescribed within 9 months of randomization).

The outcome of Aim 3 is HIV virological suppression defined as an undetectable plasma HIV viral load at the 12-month visit (a binary variable). Another secondary outcome for Aim 3 will be HIV virological suppression at the 6-month visit (a binary variable).

Another secondary outcome is INH preventive therapy discontinuation due to grade 3 or grade 4 hepatoxicity at any time while receiving INH (a binary outcome). Grade 3 hepatotoxicity was defined as AST or ALT >=5x ULN or based on symptoms that met pre-specified Grade 3 criteria for hepatotoxicity. Grade 4 hepatotoxicity was defined as AST or ALT >=10x ULN or based on symptoms that met pre-specified Grade 4 criteria for hepatotoxicity.

Finally, active TB during the 12 months of follow-up (a binary outcome) is also a secondary outcome. Active TB is defined as confirmed (if Xpert MTB/RIF assay positive) or suspected (based on chest x-ray findings or response to anti-TB treatment in a symptomatic and Xpert assay negative person).

**Descriptive statistics**

We will report the distribution of baseline covariates overall and by randomization arm. For continuous variables, we will provide the median, mean, standard deviation, 0th, 25th, 50th, 75th, and 100th percentiles. For categorical variables, we will provide frequencies and proportions. Differences between arms will not be evaluated based on statistical significance.

Participants who met eligibility criteria and enrolled in the study will be compared with eligible participants who declined enrollment by variables collected during screening, using the 2 independent samples t-test and Fisher's exact test.

**Analyses of the primary outcomes**

Primary analysis - factorial analysis. For each primary outcome, we will use a multivariable logistic regression model to estimate the effect of the relevant intervention while adjusting 1) for the other intervention and 2) the stratification randomization factors (gender and study site). We will then use the parameter estimates from this logistic regression model to predict the probability of the outcome under the intervention versus no intervention and obtain the adjusted absolute risk (proportions with outcome value equal to 1) for intervention and no intervention, the risk difference (primary treatment effect measurement), and risk ratio with 95% confidence intervals and p-values.

The trial focuses on two distinct domains: level of drinking - with a drinking outcome and an intervention consisting of economic incentives conditional on drinking- and level of adherence - with an INH adherence outcome and an intervention consisting of economic incentives conditional on INH adherence. Each domain is conceptually different and deserves to be tested at their own 5% two-sided level of significance.

<u>Estimation of adjusted absolute risks, risk difference and risk ratios for the drinking outcome</u>

Let $\pi_i$ be the probability of the individual $i$ to experience the alcohol outcome, $alcohol\ int_i$ a binary indicator for whether the individual *i* was randomized to receive the alcohol intervention (0: arms 1 or 3; 1: arms 2 or 4), $adherence\ int_i$ a binary indicator for whether the individual $i$ was randomized to receive the INH adherence intervention (0: arms 1 or 2; 1: arms 3 or 4), and $z_i$ the value of the randomization stratification factor. The model will be parametrized as:

$$logit(\pi_i) = \beta_0 + \beta_1 Alcohol\ Int_i + \beta_2\ Adherence\ Int_i +\ \beta_3\ z_i$$

Given that our goal is to estimate the effect of the alcohol intervention on the alcohol outcome, we will use the parameter estimates from this logistic regression model to estimate the predicted probability of the alcohol outcome $\pi_i$ for each individual $i$ had they received the alcohol intervention:

$$\hat{\pi}_i(Alcohol\ Int = 1) = \frac{\exp(\hat{\beta}_0 + \hat{\beta}_1 + \hat{\beta}_2\ Adherence\ Int_i +\ \hat{\beta}_3\ z_i)}{1 + \exp(\hat{\beta}_0 + \hat{\beta}_1 + \hat{\beta}_2\ Adherence\ Int_i +\ \hat{\beta}_3\ z_i)}$$

and had they not received the alcohol intervention:

$$\hat{\pi}_i(Alcohol\ Int = 0) = \frac{\exp(\hat{\beta}_0 + \hat{\beta}_2\ Adherence\ Int_i +\ \hat{\beta}_3\ z_i)}{1 + \exp(\hat{\beta}_0 + \hat{\beta}_2\ Adherence\ Int_i +\ \hat{\beta}_3\ z_i)}$$

The adjusted risks of the alcohol outcome under the alcohol intervention and under no alcohol intervention will computed as:

Risk $(Alcohol\ Int = 1) = \frac{1}{n}\sum_{i=1}^{n}(\hat{\pi}_i(Alcohol\ Int = 1)$

Risk $(Alcohol\ Int = 0) = \frac{1}{n}\sum_{i=1}^{n}(\hat{\pi}_i(Alcohol\ Int = 0)$

The risk difference and risk ratio will be computed as:

$$RD = \frac{1}{n}\sum_{i=1}^{n}(\hat{\pi}_i(Alcohol\ Int = 1) - \hat{\pi}_i(Alcohol\ Int = 0)$$

$$RR = \frac{1}{n}\sum_{i=1}^{n}(\hat{\pi}_i(Alcohol\ Int = 1)/\frac{1}{n}\sum_{i=1}^{n}(\hat{\pi}_i(Alcohol\ Int = 0)$$

Standard errors and confidence intervals will be derived using the delta method using the *adjrr* command in Stata [1].

<u>Estimation of adjusted absolute risks, risk difference and risk ratios for the INH adherence outcome</u>

Let $\pi_i$ be the probability of the individual $i$ to experience the INH adherence outcome, $adherence\ int_i$ a binary indicator for whether the individual $i$ was randomized to receive the INH adherence intervention (0: arms 1 or 2; 1: arms 3 or 4), $alcohol\ int_i$ a binary indicator for whether the individual *i* was randomized to receive the alcohol intervention (0: arms 1 or 3; 1: arms 2 or 4), and $z_i$ the value of the randomization stratification factor. The model will be parametrized as:

$$logit(\pi_i) = \beta_0 + \beta_1 Adherence\ Int_i + \beta_2\ Alcohol\ Int_i + \ \beta_3\ z_i$$

Given that our goal is to estimate the effect of the adherence intervention on the INH adherence outcome, we will use the parameter estimates from this logistic regression model to estimate the predicted probability of the INH adherence outcome, $\pi_i$, for each individual $i$ had they received the INH adherence intervention:

$$\hat{\pi}_i(Adherence\ Int = 1) = \frac{\exp\ (\hat{\beta}_0 + \hat{\beta}_1 + \hat{\beta}_2\ Alcohol\ Int_i + \ \hat{\beta}_3\ z_i)}{1 + \exp\ (\hat{\beta}_0 + \hat{\beta}_1 + \hat{\beta}_2\ Alcohol\ Int_i + \ \hat{\beta}_3\ z_i)}$$

and had they not received the INH adherence intervention:

$$\hat{\pi}_i(Adherence\ Int = 0) = \frac{\exp\ (\hat{\beta}_0 + \hat{\beta}_2\ Alcohol\ Int_i + \ \hat{\beta}_3\ z_i)}{1 + \exp\ (\hat{\beta}_0 + \hat{\beta}_2\ Alcohol\ Int_i + \ \hat{\beta}_3\ z_i)}$$

The adjusted risks of the INH adherence outcome under the INH adherence intervention and under no INH adherence intervention will computed as:

Risk $(Adherence\ Int = 1) = \frac{1}{n}\sum_{i=1}^{n}(\hat{\pi}_i(Adherence\ Int = 1)$

Risk $(Adherence\ Int = 0) = \frac{1}{n}\sum_{i=1}^{n}(\hat{\pi}_i(Adherence\ Int = 0)$

The risk difference and risk ratio will be computed as:

$$RD = \frac{1}{n}\sum_{i=1}^{n}(\hat{\pi}_i(Adherence\ Int = 1) - \hat{\pi}_i(Adherence\ Int = 0)$$

$$RR = \frac{1}{n}\sum_{i=1}^{n}(\hat{\pi}_i(Adherence\ Int = 1) / \frac{1}{n}\sum_{i=1}^{n}(\hat{\pi}_i(Adherence\ Int = 0)$$

Standard errors and confidence intervals will be derived using the delta method using the *adjrr* command in Stata [1].

Analysis of interaction between the two interventions. The DIPT trial was designed as a 2x2 factorial trial under the assumption of no interaction between the alcohol and INH adherence interventions [2]. The rationale for this assumption is that 1) if an interaction between the alcohol and INH adherence interventions exists it is expected to be synergistic, i.e., arm 4 having better outcomes than arm 3 or arm 2, and 2) only a strong synergy between the two interventions would invalidate our study design, which is not expected.

The study is not powered to detect an interaction between the two interventions as significant. However, because assessing the size of the interaction term in factorial designs is often recommended [3, 4], for each primary outcome we will test for an interaction by adding an interaction term between the two interventions in the logistic regression models defined above for the primary outcomes. More specifically, for each outcome, we will use a logistic regression model to estimate the effect of the relevant intervention including an interaction term between the two interventions and adjusting for the other intervention, and the stratification randomization factors. We will use a likelihood ratio test to test for the interaction at a 5% level of significance.

If for a given outcome no significant interaction between the two interventions is found, then no further action is required. If for a give outcome a significant interaction is found, then we will report the effect of the intervention stratified by levels of the other intervention. More specifically, if there is significant interaction between the alcohol and INH adherence intervention on the alcohol outcome, we will report two effects of the alcohol intervention, one for each level of the adherence intervention: 1) the effect of Arm 2 (alcohol intervention only) versus Arm 1 (control) and 2) the effect of Arm 4 (alcohol and INH adherence intervention) versus Arm 3 (INH adherence intervention only). Similarly, if there is if there is a significant interaction between the alcohol and INH adherence intervention on the INH adherence outcome, we will report two effects of the INH adherence intervention, one for each level of the alcohol intervention: 1) the effect of Arm 3 (INH adherence intervention only) versus Arm 1 (control) and 2) the effect of Arm 4 (alcohol and INH adherence intervention) versus Arm 2 (alcohol intervention only). Several conflicting viewpoints are expressed in the literature regarding the circumstances in which a multiple-testing correction should be used [5, 6]. For these analyses we choose not to adjust for multiplicity and use a 5% level of significance to test the effect of the intervention, while explicitly reporting a priori the comparisons that will be made.

Missing data: The primary statistical analyses will consist of a complete case analysis of individuals with non-missing outcomes. However, participants who miss the 3-month or/and 6-month visit will have missing data on the drinking outcome, which requires PEth measurements and AUDIT-C at both 3 and 6 months. To examine potential missing data mechanisms, we will compare the characteristics of the individuals who had complete data and missing data on the outcome. To examine the robustness of our findings, we will conduct a sensitivity analysis conservatively assuming that individuals who missed a study visit at 3 and/or 6 months were drinking heavily (alcohol outcome). This assumption is reasonable in this population of heavy drinkers. Finally, if it is reasonable to assume that data are missing at random and the proportion of missing data is more than 10%, we will consider using inverse probability weighting to impute the missing outcomes.

**Analyses of the secondary outcomes**

For the hepatoxicity outcome, we will use a logistic regression model including indicator variables to represent the study arms and, to improve efficiency, the randomization stratification factors (gender and study site). We will conduct 3 pairwise comparisons between each intervention (Arms 2, 3, and 4) versus the control (Arm 1).

For all other binary secondary outcomes, we will use the same statistical approach we described for the primary outcomes including a factorial analysis and a test for interaction between the interventions.

If the statistical analyses indicate an effect of the alcohol intervention versus no alcohol intervention on HIV virological suppression, we will then conduct a mediation analysis to explore the role of alcohol drinking as a potential mediator that may drive the intervention to improve HIV virologic suppression. Similarly, if the statistical analyses indicate an effect of the INH adherence intervention versus no INH adherence intervention on HIV virological suppression, we will then conduct a mediation analysis to explore the role of INH adherence as a potential mediator that may drive the intervention to improve HIV virologic suppression. More specifically, we will use methods for mediation analysis to derive the direct and indirect effects of the incentives for the primary outcomes while allowing for exposure-mediator interactions [7, 8]. We will specify a priori the list of mediator-outcome confounders.

For all continuous secondary outcomes, we will fit linear regression models. We will explore transformations for variables that are not normally distributed. If an appropriate transformation is not identified, median regression will be used. For the continuous alcohol outcomes, we will also adjust the models for the baseline drinking variables (analysis of covariance).

For all secondary outcomes, we will declare statistical significance if the p-value is < 0.05.

<u>Missing data.</u> For all secondary analyses we will conduct a complete case analysis.

**Exploratory analyses**

Exploratory outcomes. Exploratory outcomes for Aim 1 are the following continuous measures of drinking: 1) self-reported number of drinking days in the prior 30 days, 2) number of heavy drinking days (defined as ≥4 drinks/occasion and ≥5 drinks/occasion for women and men, respectively) in the prior 14 days, and 3) PEth as a continuous variable measured at other time points.

An exploratory outcome for Aim 2 will be drug concentration (ng/mg) in a subset of small hair samples. Hair levels of INH measure long-term and cumulative exposure to INH will be used to determine if short-term increases in adherence translate to sustained changes in behavior.

Intervention-by-time-interaction. For Aims 1 and 2, we will check for the presence of intervention-by-time interaction by conducting separate analyses at months 3 and 6. We will use generalized estimating equations with an independent working correlation matrix for binary outcomes and random effect models for continuous outcomes.

Impact of COVID-related disruptions. Recruitment started on April 16 2018 and study activities were paused in March 19, 2020 due to a lockdown enforced by the Ugandan government in response to the COVID-19 pandemic. During this time, all research activities were stopped and participants on INH were contacted and instructed to stop taking their INH pills immediately, as there were no means to effectively monitor for liver toxicity. Study visits resumed on May 26, 2020 and enrollment resumed on June 16, 2020. We will conduct several descriptive statistics to examine to what extent the lockdown and the pandemic might have impacted the generalizability of the trial's findings. First, we will group the participants based on whether their scheduled INH completion date was before versus on or after March 19 2020 and will compare their baseline characteristics including measures of drinking. Second, we will perform an analysis to explore whether completion of INH before or after pandemic interruption acts as potential effect modifier of the incentive interventions.

Subgroup analyses. We will perform analyses to explore potential effect modifiers of the incentive interventions. For Aims 1 and 2, the potential moderators of interest are: sex, baseline level of alcohol use, time and risk preferences, and readiness to change . For each aim, we will fit separate models including 2-way interactions between randomization group and each potential effect moderator, as well as a model testing the interaction between the two interventions. If an interaction is significant, subsequent stratified analyses will be conducted to explore the effect of the economic incentive interventions by categories of the moderator.